\input{aipcheck}
\documentclass[
    ,final            
  ]
  {aipproc}

\layoutstyle{6x9}

\usepackage{amsmath}
\usepackage{amssymb}

\begin{document}

\title{Tentative observation of a gamma-ray line\\ at the Fermi-LAT}


\author{Christoph Weniger}{
  address={Max-Planck-Institut f\"ur Physik, F\"ohringer Ring 6, 80805
  M\"unchen, Germany}, email={weniger@mppmu.mpg.de}
}

\begin{abstract}
  Using 43 months of public gamma-ray data from the Fermi Large Area Telescope,
  we find in regions close to the Galactic center at energies of 130 GeV a
  4.6$\sigma$ excess that is not inconsistent with a gamma-ray line from dark
  matter annihilation.  When taking into account the look-elsewhere effect, the
  significance of the observed signature is 3.2$\sigma$. If interpreted in terms
  of dark matter particles annihilating into a photon pair, the observations
  imply a partial annihilation cross-section of about $10^{-27}$ cm$^3$ s$^{-1}$
  and a dark matter mass around 130 GeV. We will review aspects of the
  statistical analysis and comment on possible instrumental
  indications.
\end{abstract}

\maketitle

\section{Introduction}
Searches for signatures from dark matter (DM) annihilation in the gamma-ray data
are plagued by the question of how to disentangle astrophysical foregrounds
from an actual DM signal. Gamma-ray lines, as well as the sharp pronounced
features coming from final-state radiation and virtual internal
Bremsstrahlung, are long known to provide smoking gun signatures for DM
annihilation, as they would clearly stand out of the continuous astrophysical
background flux.

The recent discovery of line-like features around 130 GeV in the Fermi Large
Area Telescope (LAT) data by Bringmann et al.~\cite{Bringmann:2012vr} and
Weniger~\cite{Weniger:2012tx}, and its subsequent confirmation by Tempel et
al.~\cite{Tempel:2012ey} and Su \& Finkbeiner~\cite{Su:2012ft}, has initiated
a torrent of activities trying to explain this signature alternatively as a
signal for DM annihilation, exotic pulsar winds~\cite{Aharonian:2012cs}, or
studying the possibility that it could be due to an instrumental
effect~\cite{Finkbeiner:2012ez, Hektor:2012ev} (for a recent
review on indirect DM searches via gamma rays, with emphasize the 130 GeV
feature and a complete list of relevant references, see
Ref.~\cite{Bringmann:2012ez}).

In this proceedings contribution, we shortly review the methods that were
adopted in Refs.~\cite{Bringmann:2012vr, Weniger:2012tx} to reveal the line
structure, comment on possible instrumental indications (for a detailed
discussion see Ref.~\cite{Finkbeiner:2012ez}), and we close with a brief
outlook.

\section{Gamma-ray lines and Fermi-LAT}

Almost all existing searches for gamma-ray lines from WIMP annihilation are
based on a dedicated analysis of the gamma-ray energy spectra measured in
regions-of-interest (ROIs) with large S/N for DM signals. The basic philosophy
is to look for line-like features on top of a smooth background spectrum. In
order to trade the uncertainties in the background flux for statistical
errors, the spectral analysis is typically confined to a small energy range
around the line energy of interest. The smooth background spectra can then be
approximated -- at first order -- by a single power-law.\medskip

\begin{figure}
  \includegraphics[width=.49\textwidth]{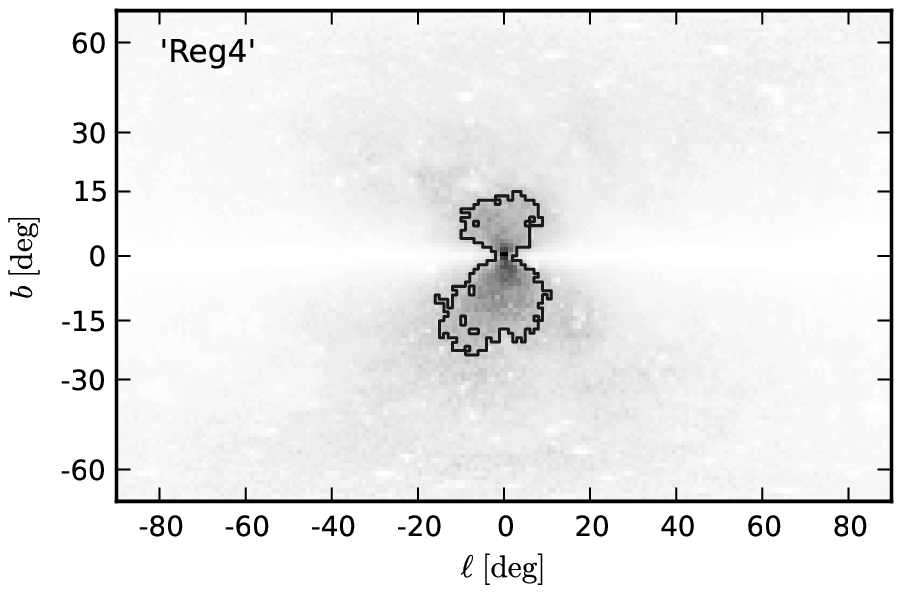}
  \includegraphics[width=.52\textwidth]{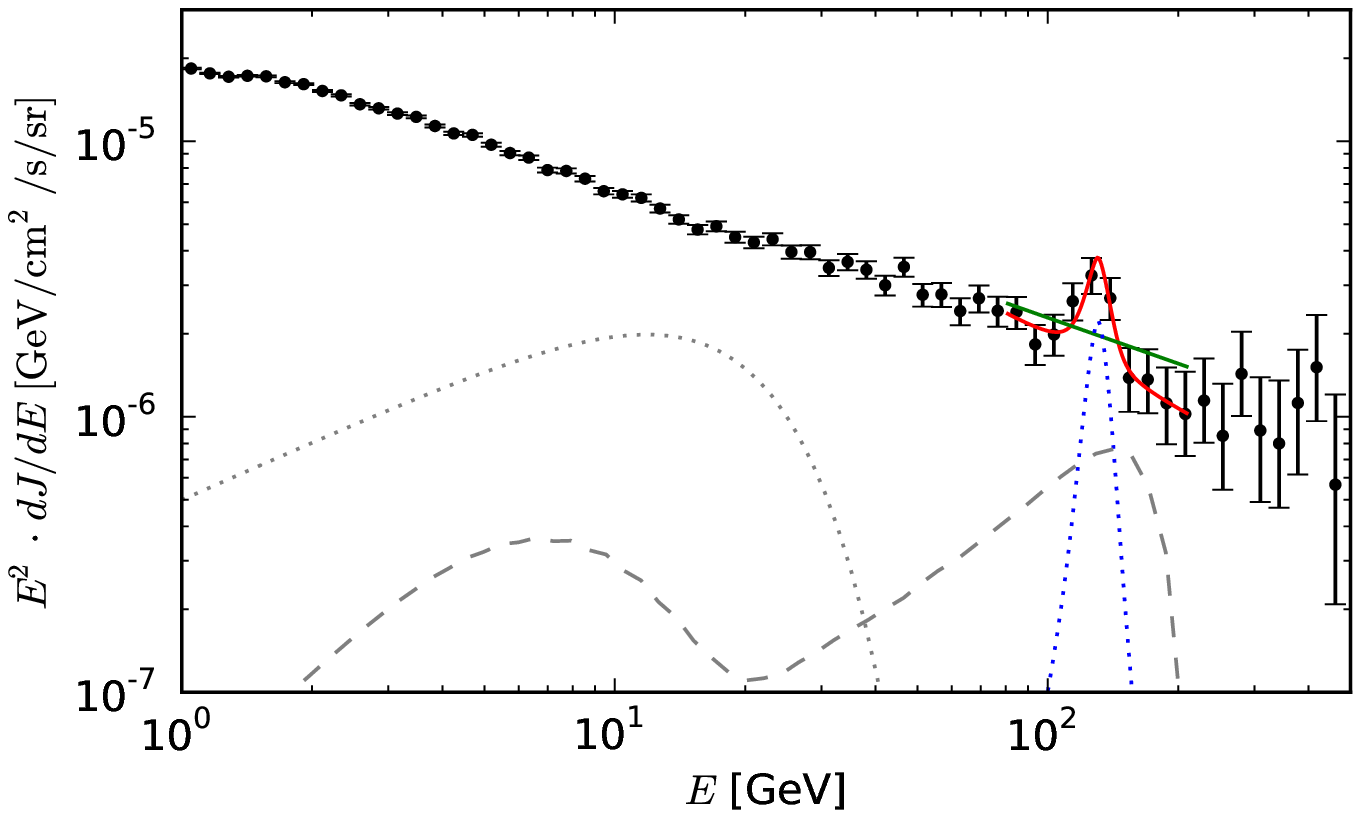}
  \caption{\emph{Left panel:} Target region Reg4 from \cite{Weniger:2012tx},
  optimized for large S/N in case of slightly contracted profile. \emph{Right
  panel:} Gamma-ray flux measured within that region by Fermi-LAT. An excess
  of events around 130 GeV is clearly visible in the data. We show the fits to
  the data in the energy range 80--210 GeV (see Ref.~\cite{Weniger:2012tx} for
  details). For direct comparison we show a very hard spectrum with
  super-exponential cut-off (left dotted line; $\sim E^{-1.3}\exp[-(E/20{\ \rm
  GeV})^2]$) and the ICS emission from monoenergetic 230 GeV electrons at the
  GC (dashed), both with arbitrary normalization.}
  \label{fig:regSpec}
\end{figure}

The main novelty of Refs.~\cite{Bringmann:2012vr, Weniger:2012tx} (besides
including all available Fermi-LAT data) is the use of an adaptive method to
find ROIs optimized for different profiles of the Galactic DM halo. One of
these regions (for a slightly contracted profile) is shown in the left panel
of Fig.~\ref{fig:regSpec} by the black line~\cite{Weniger:2012tx}.  The
differential flux measured from this ROI is plotted in the right panel; it
exhibits a surprisingly clear and sharp excess of events around $130$~GeV.
The solid lines to the right represent a power-law only (power-law + line
signal) fit to the data, restricted to the range 80--210 GeV. The local
significance for the presence of a line signal is
$4.6\sigma$~\cite{Weniger:2012tx} (even higher significances were found in the
template analysis of Ref.~\cite{Su:2012ft}). The gray dotted line at the left
shows for comparison a hard spectrum with super-exponential cutoff; the gray
dashed line shows the inverse Compton scattering (ICS) radiation from a
mono-energetic electron population at the Galactic center (GC). None of the
spectra is sharp enough to fit the 130 GeV excess.

\bigskip

In order to check whether the observed excess is due to a mis-calibrated
effective area or energy reconstruction we analysed test samples with a very
low S/N for a DM line signal, like regions along the Galactic disk or the
Earth limb.

\begin{figure}
  \includegraphics[width=.5\textwidth]{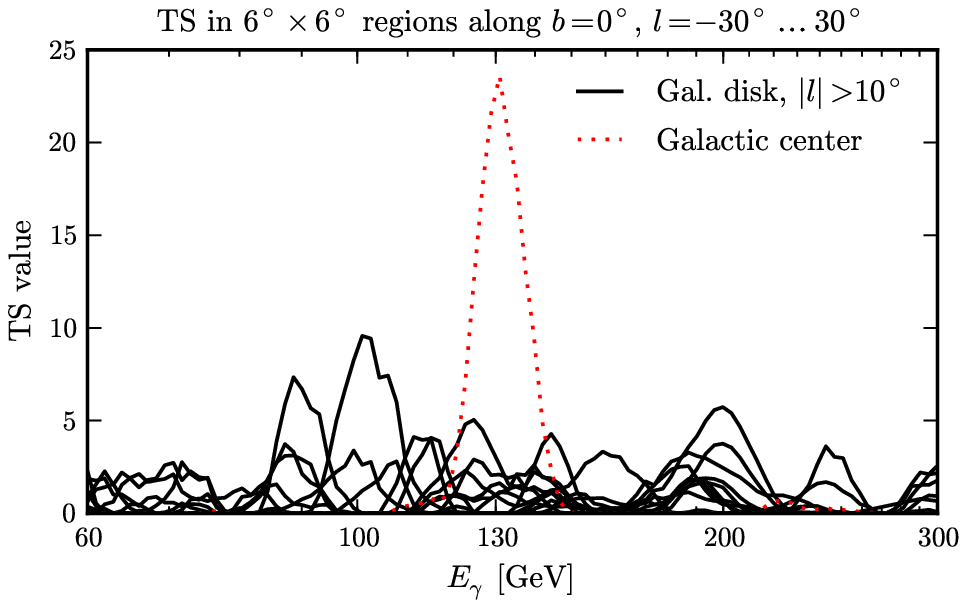}
  \includegraphics[width=.5\textwidth]{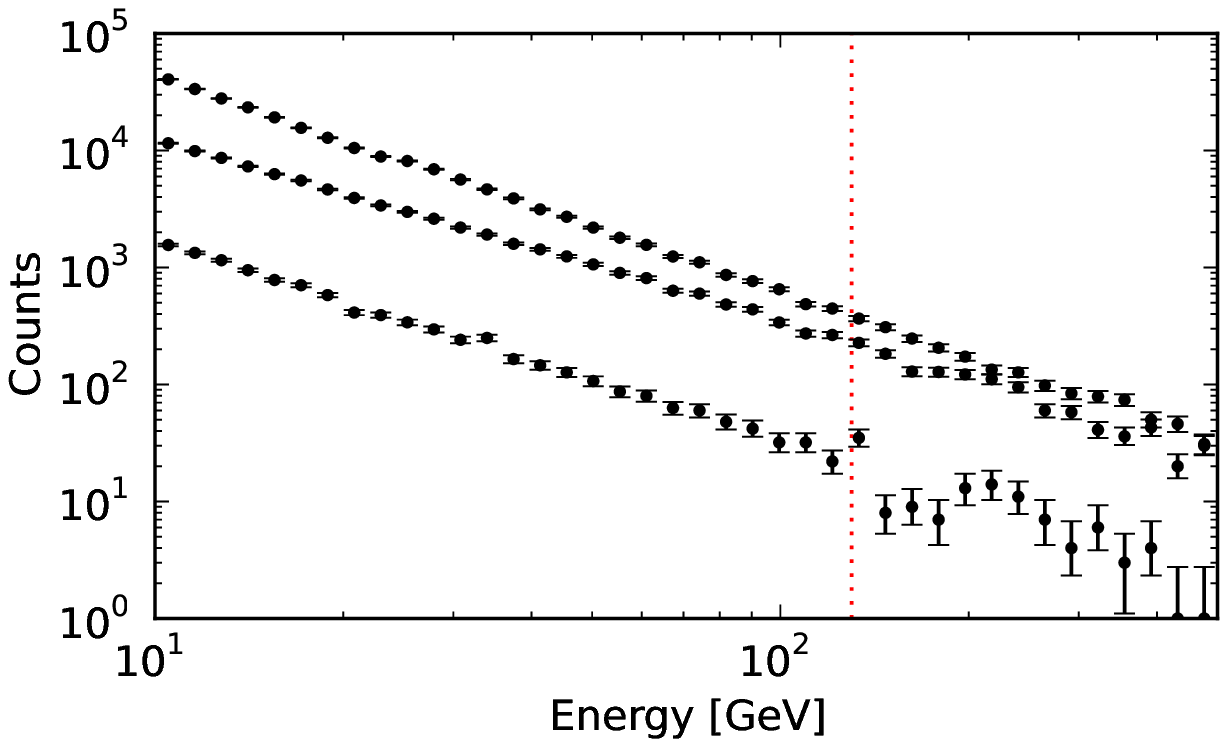}
  \caption{\emph{Left panel:} TS value in different $6^\circ\times6^\circ$
  regions along the Galactic plane, centered on $|\ell|=0^\circ$ (dotted line)
  and $|\ell|=12^\circ, 15^\circ, \dots, 30^\circ$ (solid lines). \emph{Right
  panel:} from top to bottom this plot shows the Earth limb events
  ($Z>110^\circ$), all events with incidence angles in the range
  $\theta=30^\circ\dots45^\circ$, and Earth limb events with the same
  incidence angle range ($Z>110^\circ$ \&\& $\theta=30^\circ\dots45^\circ$, see also
  'suspicious Earth limb line' events in Ref.~\cite{Finkbeiner:2012ez}.)}
  \label{fig:TStest}
\end{figure}

In the left panel of Fig.~\ref{fig:TStest}, we show the TS values found when
searching for lines at different energies in different regions along the
Galactic plane. The black lines correspond to partially overlapping
$6^\circ\times6^\circ$ regions away from the GC; the red line corresponds to a
region of same size but centered on the GC. The only significant line
signature is the one at the GC around 130 GeV. Fits are performed as in
Ref.~\cite{Weniger:2012tx} for SOURCE class events.

Earth limb photons, which stem from cosmic rays interacting with the Earth
atmosphere, provide a smooth reference spectrum for systematic checks.  In the
right panel of Fig.~\ref{fig:TStest}, the bottom data points shows Earth limb
photons (with zenith angles $Z>110^\circ$) that hit the LAT at incidence
angles in the range $30^\circ\lesssim\theta\lesssim45^\circ$. A significant
line-like feature or spectral distortion around 130 GeV appears, with a formal
local significance well above $3\sigma$~\cite{Finkbeiner:2012ez}.  However,
when considering \emph{all} events with incidence angles
$30^\circ\lesssim\theta\lesssim45^\circ$ but any zenith angle $Z$ (central
data points), or \emph{all} Earth limb events ($Z>110^\circ$) with any
incidence angle $\theta$ (top data points), the structure disappears. Note
that although an instrumental effect can depend on $\theta$, it would be
surprising if it depends on $Z$; a feature in the physical flux can depend on
$Z$ but not on $\theta$.  Clearly, more data from the Earth limb is needed to
check whether the feature in the low incidence angle Earth limb data is a
statistical fluke or points to a serious problems with the processing of
$>100$~GeV PASS 7 events (for more details see Ref.~\cite{Finkbeiner:2012ez}).

\begin{figure}
  \includegraphics[width=.49\textwidth]{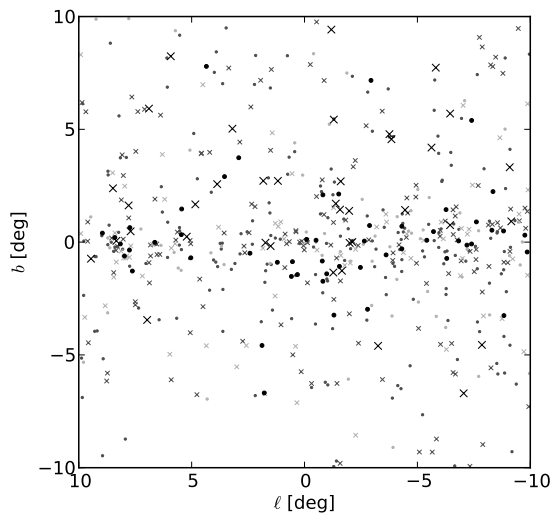}
  \caption{Spatial distribution of line events close to 130 GeV (CLEAN events
  from 4 Aug 2008 to 5 Sep 2012). Black dots (crosses) correspond to front
  (back) converting events in the range 120--140~GeV. The dark (light) gray
  dots/crosses show the sidebands 70--110~GeV (150--300~GeV) for comparison.
  The events around 130 GeV are clustered slightly west from the GC.}
  \label{fig:countmap}
\end{figure}

In Ref.~\cite{Su:2012ft}, it was found that the 130 GeV feature shown in
Fig.~\ref{fig:regSpec} is not exactly centered on the GC; instead, a
$\sim1.5^\circ$ shift to the west of the GC is observed. The events
contributing to the 130 GeV excess are shown in black in
Fig.~\ref{fig:countmap} as function of Galactic longitude $\ell$ and latitude
$b$; the adopted energy range is $120$--$140$~GeV.  Already this plot suggests
that the 130 GeV excess is not centered on the GC, although considerably more
data is required to make robust statements about its distribution.  For
comparison, we also show the $70$--$110$~GeV and 150--300 GeV sidebands in
dark and light gray, respectively.

\section{Conclusions}
The observation of a gamma-ray line close to the GC would be an impressive
smoking gun signature for annihilation of DM in the Universe. In the Fermi-LAT
data, we find strong indications for such an excess at around 130 GeV. The
formal significance is high, making it unlikely to be a statistical fluke,
although a final confirmation requires more data to reproduce the signal
without trials. Right now, the main concern is whether the PASS 7 processing
of LAT events is reliable at high energies, where the low statistics still
limits a sufficient in-flight study of instrumental effects.  A first
indication for such an instrumental effect might be the observation of a
line-feature at 130 GeV in a subset of the Earth limb data, although this
feature is not present in other more orthodox test samples. In any case, it
would be difficult to understand how such feature could map onto the GC.
However, even after trials, the probability for the Earth limb line to be
produced by chance is already uncomfortably low and about $\sim1\%$.  Clearly,
more data is urgently required, to clarify whether this is a statistical fluke
or a real effect.

\begin{theacknowledgments}
  The author thanks the organizers of the \emph{5th International Symposium on
  High Energy Gamma-Ray Astronomy (Gamma2012)}, Heidelberg July 9--13, for a
  very interesting and stimulating conference, and acknowledges partial
  support from the European 1231 Union FP7 ITN INVISIBLES (Marie Curie
  Actions, PITN-GA-2011-289442).
\end{theacknowledgments}

\bibliographystyle{aipproc}
\bibliography{}

\begin{thebibliography}{8}
\expandafter\ifx\csname natexlab\endcsname\relax\def\natexlab#1{#1}\fi
\providecommand{\enquote}[1]{``#1''}
\expandafter\ifx\csname url\endcsname\relax
  \def\url#1{\texttt{#1}}\fi
\expandafter\ifx\csname urlprefix\endcsname\relax\def\urlprefix{URL }\fi
\providecommand{\eprint}[2][]{\url{#2}}

\bibitem[Bringmann et~al.(2012)]{Bringmann:2012vr}
T.~Bringmann, X.~Huang, A.~Ibarra, S.~Vogl, and C.~Weniger, \emph{JCAP}
  \textbf{1207}, 054 (2012), \eprint{1203.1312}.

\bibitem[Weniger(2012)]{Weniger:2012tx}
C.~Weniger, \emph{JCAP} \textbf{1208}, 007 (2012), \eprint{1204.2797}.

\bibitem[Tempel et~al.(2012)]{Tempel:2012ey}
E.~Tempel, A.~Hektor, and M.~Raidal  (2012), \eprint{1205.1045}.

\bibitem[Su and Finkbeiner(2012)]{Su:2012ft}
M.~Su, and D.~P. Finkbeiner  (2012), \eprint{1206.1616}.

\bibitem[Aharonian et~al.(2012)]{Aharonian:2012cs}
F.~Aharonian, D.~Khangulyan, and D.~Malyshev  (2012), \eprint{1207.0458}.

\bibitem[Finkbeiner et~al.(2012)]{Finkbeiner:2012ez}
D.~P. Finkbeiner, M.~Su, and C.~Weniger  (2012), \eprint{1209.4562}.

\bibitem[Hektor et~al.(2012)]{Hektor:2012ev}
A.~Hektor, M.~Raidal, and E.~Tempel  (2012), \eprint{1209.4548}.

\bibitem[Bringmann and Weniger(2012)]{Bringmann:2012ez}
T.~Bringmann, and C.~Weniger  (2012), \eprint{1208.5481}.

\end{thebibliography}

\end{document}